\UseRawInputEncoding
\documentclass[journal]{IEEEtran}
\usepackage{cite}
\usepackage[dvips]{graphicx}
\DeclareGraphicsExtensions{.eps}
\usepackage[cmex10]{amsmath}
\usepackage{array}
\usepackage{mdwmath}
\usepackage{mdwtab}
\usepackage{caption}
\usepackage{amsfonts}
\usepackage{url}
\usepackage{balance}
\usepackage{algorithm}
\usepackage{algorithmic}
\usepackage{lmodern}
\usepackage[tight,footnotesize]{subfigure}
\usepackage[caption=false,font=footnotesize]{subfig}

\ifCLASSINFOpdf
\else
\fi

\begin{document}
	%
	\title{On Routing, Wavelength, Network Coding Assignment and Protection Configuration Problem in Optical-processing-enabled Networks}
	%
	%
	%
	
	\author{Dao Thanh Hai 
	}

	\markboth{Journal of \LaTeX\ Class Files,~Vol.~xxx, No.~xxx, August~2017}%
	{Shell \MakeLowercase{\textit{et al.}}: Bare Demo of IEEEtran.cls for IEEE Journals}

	\maketitle
	
	\begin{abstract}
		In optical-processing-enabled network, transitional lightpaths crossing the same node could be optically encoded to each other to achieve greater spectral efficiency. In this context, we present a new research problem, entitled, routing, wavelength, network coding assignment and protection configuration (RWNCA-PC) arisen in exploiting photonic network coding (NC) for dedicated path protection in wavelength division multiplexing (WDM) networks with an extra degree of freedom in the selection of protection triggering mechanism, that is, network-side and client-side, tailoring to each connection. In order to maximize the NC benefits, we thus provide a weighted multi-objective optimization model for solving RWNCA-PC problem so as to minimize the wavelength count as the strictly prioritized goal and the redundant resources measured by the number of client-side connections as the secondary objective. Numerical results on the realistic COST239 network reveal that a saving of up to $25\%$ wavelength resources could be achieved thanks to the optimal use of NC compared to the non-coding designs and among coding-aware designs, the use of mixed protection configurations would be spectrally more efficient than the design with only network-side protection scheme. Our proposal yields the highest spectrum efficiency compared to all reference designs and moreover, features an average saving of more than $40\%$ transponder count compared with its single objective counterpart. 	
	\end{abstract}
	
	\begin{IEEEkeywords}
		Optical-processing-enabled Network, Optical-processing Networking, Optical-processing-enabled Node, Protection Configuration, Multi-objective Optimization, Photonic XOR, Network Coding, Routing, Wavelength Assignment, Integer Linear Programming
	\end{IEEEkeywords}

	%
	\IEEEpeerreviewmaketitle

	\section{Introduction}
	\label{sec1}
	Network traffic will face a dramatic growth over the next coming years driven by the widespread adoption of 5G technologies and the rise of immersive and interactive services such as holographic communications and tele-haptics \cite{Cisco20}. Such an explosive increase is posing unprecedented challenges to transport networks, causing a critical concern on the crunch of fiber capacity. In addressing this major problem, technological and architectural solutions have been constantly proposed to accommodate traffic in a greater efficiency. With regard to network architecture, optical networks have been progressively transitioning from the optical-electrical-optical (O-E-O) mode to transparent (all-optical) one and in doing so, a significant amount of intermediate electronic processing equipments have been eliminated, resulting in both capital and operational savings \cite{all-optical, hai_icact}. From technological sides, conventional paths\textemdash for many years\textemdash have been focusing on improving spectral efficiency by higher-order modulation formats, super-channel and flexible optical networks both in spectral and spatial dimensions \cite{hai_wiley, spatial1, spatial3}. It has been clear that almost all known physical dimensions for multiplexing have been proposed and exploited and this necessitates for a departure from traditional schemes to seek for innovative techniques to further enhance network efficiency. \\ 
	
	In an information-driven era, securing the continuity of information flows has been of critical importance. The crash of Amazon networks and recently Facebook in just a few hours causing a tremendous financial losses, along with other serious repercussions due to business interruptions for millions of customers, has provoked growing alarm on protecting the critical infrastructure and underlined the rising need for providing fast recovery services \cite{Amazon}. In practice, the 1+1 dedicated path protection (DPP) has been widely deployed in such networks to offer hit-less recovery in an agile way thanks to maintaining two active paths\textemdash working and protection\textemdash between the transmitter and receiver, and thus a simple decision circuitry is merely needed at the receiver to pick up the better one \cite{hai_ps2, hai_rtuwo, hai_csndsp}. Although DPP is the fastest protection scheme available, it suffers from inefficient capacity utilization because of doubling the required resources. As the fiber capacity crunch concern is imminent, providing dedicated protection services thus appears to be increasingly challenging for network operators. \\
	
	Since its inception in \cite{NC}, the applications of network coding (NC) in communication networks have been progressing dramatically and generating abundant research efforts \cite{NC-survey}. The use of NC techniques empower intermediate nodes with processing capabilities for incoming signals rather than simply having conventional forwarding functions and such feature acts as a new \textit{degree of freedom} for boosting the network performance as the resources could be better utilized. In optical networks, NC appears to be a good match with the protection requirement since the use of NC allows mixing signals at appropriate nodes and transmitting such mixed one to the destination for resources savings. This perspective is radically different from traditional optical networking paradigm where individual signals are simply routed end-to-end. Indeed, integrating NC to dedicated protection has been receiving increasing attention, collectively noted as network coding-based protections \cite{Kamal}, as it marks an unconventional approach, challenging the long-established wisdom of trading capacity efficiency for speed recovery and vice versa. It is noticed that previously published works\textemdash on the application of NC for protection in optical networks\textemdash have been mostly focused on the operation of NC at packet level in conjunction with the optical-electrical-optical operations \cite{Kamal, Babarczi}. Yet the notable advances in photonic signal processing technologies \cite{xor3} and the efficiency of transparent architecture are tilting research efforts towards exploiting photonic NC operating at a wavelength granularity and in the paradigm of all-optical networks.  Among the first attempts was the work from \cite{all-optical-nc} and it was reinforced in \cite{transparent1} that all-optical NC could be of practical values as the authors provided a design framework for optimizing the use of NC. In \cite{hai_oft, hai_comletter, hai_springer}, we examined the utilization of photonic XOR for improving the capacity efficiency of traditional DPP and numerically quantified the wavelength efficiency improvement in practical networks. These works, however, centered on a single traditional metric of minimizing the wavelength count in order to support a given traffic set and thus, have not treated the further issue of classifying DPP and its impact on the network design with NC as well as on optimizing the configuration of DPP. In addressing and solving such critical concern, this paper goes an extra mile in investigating the combination of all-optical network coding and dedicated path protection in transparent WDM optical networks by digging deeper into the triggering mechanism options. When it comes to implementation, dedicated protection (1+1) is further classified into two types, that is, the client-side and network-side scheme whose the crucial difference is on where the protection mechanism is triggered (i.e., at the side of client or in the optical layer (network-side)) \cite{Simmons2014, hai_optik, hai_oft2}. While the network-side configuration requires less redundant resources at endpoints, its constraint of using the same wavelength for working and protection path might impact the network performance on wavelength resources. In contrast, the client-side triggering demands more redundant resources at endpoints and yet brings the potential advantage of greater wavelength efficiency thanks to the fact that the working and protection signal could be transmitted on different wavelengths, alleviating the wavelength assignment task. In addressing this compromise, we present a multi-objective design framework making use of both network-side and client-side schemes\textemdash mixed protection configuration\textemdash to optimally benefit from NC in terms of wavelength efficiency and furthermore, utilize a minimal number of client-side connections. In doing so, our proposal leverages the routing, wavelength and network coding assignment problem (RWNCA) \cite{hai_comletter} with added insights, that is, the selection of protection configuration tailored to each connection and this results in a new problem\textemdash routing, wavelength, network coding assignment and protection configuration (RWNCA-PC)\textemdash as the extension of the RWNCA. The network design with use of NC augmented by protection configuration selection is formulated in the form of a multi-objective optimization model aiming at minimizing the used wavelengths as the strictly prioritized goal and the number of client-side connections as the auxiliary objective. In reflecting the aforementioned priority, a detailed analysis on setting up weight coefficients for integrated objective function is provided. Numerical results on the realistic COST239 network and all-to-one traffic setting reveals that up to $25\%$ wavelength resources could be saved by making optimal use of NC compared to the non-coding designs and among coding-aware designs, the use of mixed protection configurations would be spectrally more efficient than the design with only network-side protection scheme. Our proposal achieves the highest spectrum efficiency compared to all reference designs and furthermore, features an average saving of more than $40\%$ transponders count compared with its single objective counterpart.  \\
	
	The remaining of the paper is structured as followed. Related works on the application of NC in optical networks are reviewed and then, our main contributions are positioned in Section II. Next, we highlight the issue of triggering mechanism for DPP and illustrate the use of photonic XOR as an efficient means to improve the protection efficiency of DPP in Section III. In Section IV, the mathematical formulation for optimal design of the routing, wavelength, network coding assignment and protection configuration is presented. The performance evaluation of our design proposition in comparison with reference designs on the realistic network topology is brought about in Section V. The paper is concluded in Section VI.

	\section{Literature Review and Our Contributions}
	\subsection{Literature Review}
	\label{sect: related}
	
	First, it has been noticed that the majority of research efforts in network coding have been dedicated to wireless networks in which the multi-path transmission nature appears to be a good match to network coding requirements \cite{NC-special}. In contrast, the applications of NC in optical networks have been receiving an inadequate attention as the wisdom from NC in wireless networks could not be directly transferred to optical networks. To mine the bandwidth benefits of NC in optical networks, it is necessary to have multiple signals that can be favorably mixed together \cite{Kamal08} and such condition often emerges from multi-cast scenarios \cite{nc-general3}, passive optical networks \cite{NC-PON, NC-PON3} and the case of disjoint paths in network protection \cite{Kamal} which have been prime targets of network coding research in optical networks. For the scope of this paper, our literature survey are focused on the use of NC for protection and fault tolerances. \\
	
	The research work from \cite{Kamal06} pioneered in fostering the use of NC for protecting mesh optical networks with the 1+N p-cycles protection. The authors proved that the blackout time in a failure scenario is less than the cycle delay. This pioneering work generated extensive follow-up works promoting the use of NC for protection in various scenarios, marking the arrival of network coding-based protection \cite{Kamal}. Extension of \cite{Kamal06} was made in \cite{Kamal2} focusing on implementing network coding protection on an overlay layer and thus, profiting from simple deployment, scalability and synchronization-free compared to the original scheme. The works from \cite{Kamal-pcycle2, Kamal-pcycle} extended the reach of NC for optical uni-cast traffic protection and revealed that it could be possible for sharing a single backup bandwidth unit over N connections while remaining the benefits of 1+1 protection. This technique was thus labeled as 1+N protection and could be exploited for protecting both single link and multiple link failures. In \cite{Kamal-node}, the resilience against node breakdown was re-modeled as the multiple link failures and NC was then made use of for improving the protection efficiency. Network coding protection was also applicable to the realm of multi-domain networks by combining 1+1 protection and dual homing in \cite{NC-multidomain} and it was unfolded that this combination could offer the resilience against both node and link failures in each sub-domain. The issue of combining bandwidth-intensive 1+1 protection with NC has been the focus of many studies since the merit of near-instantaneous recovery could be achieved with a greater capacity efficiency. In \cite{Belzner}, a network coding technique was proposed for reducing bandwidth utilization in 1+1 protection by relying on a simple network model composing of only two sources and a common receiving node. This work was, however, restricted to the traditional shortest path routing algorithms in combination with load balancing and therefore, incurred the non-optimal quality. In mitigating the solution quality issue, the authors from \cite{Abu} introduced an optimal network design framework in the form of integer quadratic programming to maximize the resource savings with NC. Rather than considering a simple model of two sources with a shared destination, the work in \cite{icc} went further by providing a generalized optimization model which could be applicable to arbitrarily multiple demands. An unified framework for designing NC-assisted dedicated protection in the WDM optical networks with opaque node architecture was comprehensively investigated in \cite{hai_access, hai_comcom2, hai_springer2}. To make the favorable environment for mixing signals, traffic splitting was proposed and studied in conjunction with network coding in \cite{TS1}. The research in \cite{all-optical-nc} laid the technological and algorithmic foundation for practical uses of NC in all-optical networks. Yet, this research missed to address the wavelength assignment issue which is critically important in designing, operating and managing transparent networks. In \cite{hai_comletter} and \cite{hai_oft}, the authors provided a framework for exploiting photonic NC to re-design the regular 1+1 protection in transparent WDM networks and then introduced the respective problem, that is, routing, wavelength and network coding assignment to optimize the NC impact. Recently, NC techniques have expanded its application into the elastic optical networks for improving the network efficiency with first research attempts from \cite{NC-EON} and further extended in \cite{NC-EON3, nc_security1, nc_security2}. Applications of photonic network coding for elastic optical networks, albeit very promising, would be much more difficult than for WDM networks. The most critical challenge is on the performing all-optical XOR operation between signals of different line-rate and/or modulation formats. Although it has been experimentally possible for such operations, they are nonetheless limited to simple modulation formats such as OOK, BPSK and DPSK. Therefore, compared to WDM optical networks, the technical readiness for incorporating photonic network coding to elastic optical networks appears to be more far-fetched. \\
	
	To the best of our survey, the issue of triggering mechanism for DPP in conjunction with the use of NC has not been addressed by prior works in literature and therefore, no attempt has been made to examine the impact of triggering configuration to the network performance and moreover, to optimize the protection configurations for maximal NC gain. Such critical gap is addressed and tackled in this paper. 
	
	\subsection{Contributions}
	In continuing and extending our previous works on the utilization of photonic network coding to optical layer protection \cite{hai_access, hai_comletter, hai_oft}, this paper explores the new insight for integrating all-optical network coding with dedicated path protection (DPP) to protect single link failures in transparent WDM networks by digging deeper into the implementation options of DPP, that is, network-side and client-side mechanism. As a major extension from previously published works \cite{hai_comletter} by means of differentiating the DPP node architecture into client-side and network-side types, we highlight an important observation for the first time. That is, on one hand, client-side DPP+NC may offer greater spectrum efficiency compared to network-side DPP+NC thanks to the more degree of freedom in assigning the wavelength for the working and protection signals. On the other hand, client-side DPP is costlier than the network-side scheme due to the need for two separate transponders in charge of working and protection signals compared to a single transponder in network-side DPP. Inspired by this observation, we pose a new research question, that is, how to achieve the highest spectrum efficiency with the minimal use of costly client-side connections. In tackling this question, we leverage the use of mixed protection configurations, that is, both client-side and network-side mechanism,  and present the RWNCA-PC problem, which goes beyond the RWNCA problem, by incorporating the optimal selection of protection configuration for each connection so that the spectrum efficiency is maximized while minimizing the number of transponders. Specifically, our contribution in this work are highlighted as following: \\
	
	\begin{itemize}
		
		\item The first main contribution is to investigate the triggering mechanism of DPP in transparent WDM optical networks, that is, either from client-side or network-side and propose an architectural model for integrating photonic XOR network coding to DPP to attain the near-instantaneous recovery at a lower cost of redundant capacity than the original DPP.  \\ 
		
		\item The second main contribution is to present a new research problem, namely, 1+1 Routing, Wavelength, Network Coding Assignment and \textit{Protection Configuration} (referred as RWNCA-PC). The newly formulated problem is indeed the generalization of the 1+1 routing, wavelength and network coding assignment as the triggering protection mechanism is taken into account and optimally adjusted for each connection. We then provide a mathematical framework for optimal solving of the RWNCA-PC with multiple objectives. The mathematical model aims to minimize the number of used wavelengths as the strictly prioritized goal and moreover, minimize the redundant resources at endpoints measured by the number of client-side connections as the secondary goal. The model is originally formulated in the form of a bi-objective integer non-linear programming and then be transformed into the prevailing type of integer linear programming with a linearization procedure. Furthermore, a comprehensive analysis on the impact of weight coefficients is presented as a guideline for weight selection. \\ 
		
		\item The another contribution is on extensive numerical evaluations on the realistic COST239 topology to draw a comparison between our multi-objective design proposal with mixed protection configuration and the reference designs including both network coding-based and non-coding ones. The comparison is centered on the traditional metric of wavelength resources and the relative redundant resources measured by the number of transponders.

	\end{itemize}
	\section{Integrating Photonic Network Coding to 1+1 Protection}
	\label{sect: Coding}
	
	\subsection{Node Architecture for 1+1 Dedicated Protection: Client-side vs. Network-side}
	
	\begin{figure}[!ht]
		\centering
		\includegraphics[width=\linewidth, height=0.2\textheight]{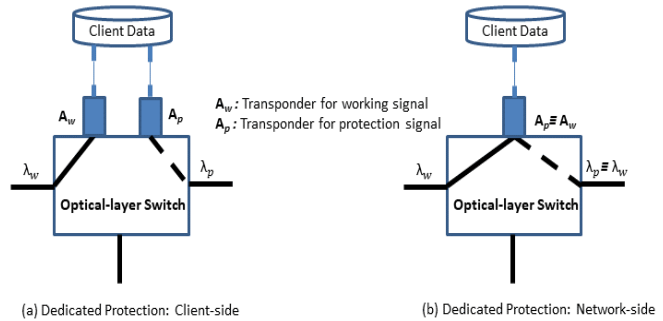}
		\caption{Client-side and Network-side Architecture for DPP}
		\label{fig:protection}
	\end{figure}
	
	1+1 dedicated optical path protection offering near-immediate recovery in case of single-link failures has been broadly adopted in practice. In implementing DPP, a further classification is required based on how two copies of signal are generated\textemdash whether they are at the side of client or in the optical layer (network-side) \cite{hai_csndsp}. Such triggering mechanisms, network-side and client-side, are translated to different ramifications on the redundant resources and network design constraints. Figures \ref{fig:protection} details the configuration of each implementation. \\ 
	
	The client-side DPP is shown in Fig. \ref{fig:protection} (a) where the client is in charge of generating two copies of the signal and the optical layer routes such signals over disjoint paths. In this configuration, thanks to having separate transponders, the wavelength for the working and protection signal of a connection could be different ($\lambda_w \ne \lambda_p$). In terms of transponder count, the client-side configuration requires \textit{two transponders} for each dedicated protection connection. Alternatively, DPP could be also implemented in the network-side mode in which the network side is responsible for having two copies of the signal routed on link-disjoint paths as in Fig. \ref{fig:protection} (b). Such network-side mechanism is clearly of lower cost thanks to the fact that a \textit{single transponder} is shared for both working and protection signal. Yet, the constraint of identical wavelength for the working and protection signal of a connection may degrade the network performance in transparent optical network designs as it would be more challenging to accommodate the wavelength assignment task. \\
	
	In recognizing the above compromise, it is therefore imperative to develop a design that could achieve the optimal spectrum efficiency while utilizing a minimal number of client-side connections and it would be more challenging as NC is exploited to leverage DPP protection.        
	
	\subsection{Network Coding for Protection}
	
\begin{figure}[!ht]
		\centering
		\includegraphics[width=\linewidth, height=0.27\textheight]{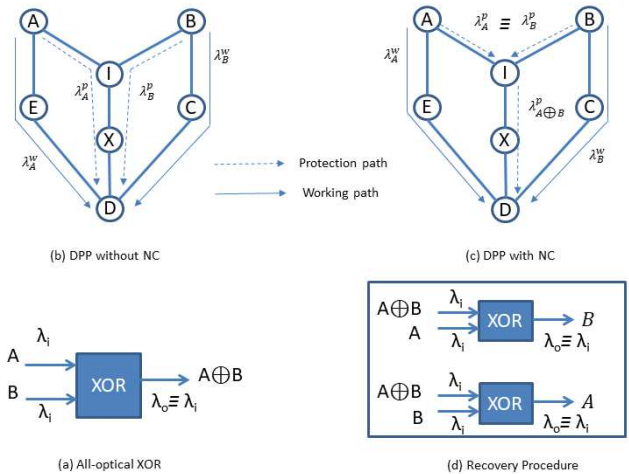}
		\caption{Network Coding for Efficient 1+1 Protection}
		\label{fig:nc}
	\end{figure}

	All-optical technologies for XOR function have been advancing remarkably and it has been demonstrated experimentally that photonic XOR could be executed at a line rate up to Tbps \cite{added3} and possibly with signals of distinctive formats \cite{xor3}. In this work, we make use of a simple photonic XOR gate where both two inputs and the output signal are modulated on an identical wavelength as illustrated in Fig. \ref{fig:nc}(a). The key merit of such scheme\textemdash XOR encoding between signals of the same wavelength\textemdash is the elimination of a probe signal and therefore, could be highly cost-efficient \cite{xor-model}. \\
	
	We present in this part an illustrative case to underline the point in making use of photonic XOR\textemdash encoding and decoding\textemdash to redesign the conventional dedicated protection in transparent WDM networks. Assuming that two traffic demands\textemdash from node $A$ to node $D$ and from node $B$ to node $D$\textemdash request dedicated protection requirements. Planning these connections in the conventional DPP setting (i.e., without NC) consists of determining the pair of link-disjoint routes and selecting proper wavelengths\textemdash working and protection wavelength\textemdash for each connection in the constraints of wavelength uniqueness and continuity. One such provisioning is presented in Fig. \ref{fig:nc}(b) and it costs \textit{two wavelengths} for the entire network as the protection routes of connection $A$ and $B$ both pass through links $IX$ and $XD$ and thus, $\lambda_A^{p} $ and $\lambda_B^{p}$ must be different. However, if NC service is provided at node $I$, protection flows of connection $A$ and $B$ could be all-optically XOR encoded\textemdash $A \oplus B$\textemdash and then transmitted transparently to the destination node $D$ on a single wavelength $\lambda_{A \oplus B}^{p} \equiv \lambda_A^{p} \equiv \lambda_B^{p}$ as illustrated in Fig. \ref{fig:nc}{(c)}. Clearly such NC-based solution paves the new opportunity for achieving greater wavelength efficiency as it simply requires \textit{one wavelength} for the whole network, representing a significant saving of $50\%$. Regarding the recovery capability in case of any single link failure, the two remaining signals are still received by the destination and the lost signal could be easily reconstructed by optical XOR operation (Fig. \ref{fig:nc}(d)).\\ 
	
	For the practical implementation of photonic network coding services in WDM optical networks, a number of technological challenges are involved. Specifically, all-optical buffer/storage and the issue of synchronization are among the most challenging things when it comes to incorporating photonic network coding into existing fiber-optics networks infrastructure. As far as all-optical buffer is involved, it could be fulfilled\textemdash in principle\textemdash by using a series of fiber delay lines onto which the optical signal can be switched \cite{added1} or by more sophisticated technology based on slow-light effects in nonlinear meta-surfaces \cite{added2}. Indeed, the analysis of the readiness of enabling technologies supporting the integration of all-optical coding to transparent optical networks has been detailed in \cite{all-optical-nc} and experimentally validated in \cite{all-optical-nc2}. It is also noticed that compared to the currently adopted forward error correction codes \cite{fec}, such XOR operation is clearly far less complex and thus expected to add trivial overhead in terms of complexity and delay. \\  
	
	To optimally profit from applying NC, the orchestration of encoding demands, encoding nodes and encoding link(s) have to be addressed in conjunction with the traditional routing and wavelength assignment. Such problem is known as the routing, wavelength and network coding assignment problem (RWNCA) \cite{hai_comletter}. When the triggering type of DPP is taken into account, the goal for minimizing the redundant resources (i.e., the number of client-side connections) is furthermore required and this results in a new problem, called, routing, wavelength, network coding assignment and protection configuration (RWNCA-PC) as the extension of the RWNCA one. In a general case, the objective is to achieve the maximal NC gain in terms of wavelength efficiency at the optimally mixed use of protection configuration and this is the focus of this study.

	\section{A Mathematical Model with an Integrated Objective for Optimal Solving of RWNCA-PC problem}
	\label{sect: math}
	This section provides a multi-objective integer linear programming formulation for the RWNCA-PC problem. Specifically, an integrated objective function is formed by adding two weighted objectives. We then present a comprehensive analysis on the impact of weight vectors to the priority of individual objectives. Inputs to the mathematical formulation are a physical topology composing of nodes and fiber links, traffic demands and the output solution determines the optimal composition including the working and protection routes for all demands, the pairs of demand for coding and respective coding nodes and coding links, and especially the protection configuration for each demand so as to minimize the wavelength count as the strictly prioritized objective and additionally minimize the client-side connections as the auxiliary goal. \\
	
	\begin{footnotesize}
		\noindent{Inputs:}
		
		\begin{itemize}
			\item $G(V,E)$: A graph represents the physical topology consisting of $|V|$ nodes and $|E|$ fiber links. Each link $e \in E$ is identified by its beginning node $s(e)$ and its ending node $r(e)$. 
			\item $D$: A set of traffic demands, indexed by $d$. Each traffic demand $d \in D$ is identified by its origin $s(d)$ and destination $r(d)$ respectively and all demands request \textit{one traffic unit of wavelength capacity}
			\item $W$: A set of available wavelengths on each fiber link, indexed by $w$. The capacity of each link measured by the number of wavelength is $|W|$
			
		\end{itemize}

		\noindent{Variables:}
		
		\begin{itemize}
			\item $x_{e, w}^{d} \in \{0,1\} $: equals 1 if demand $d$ uses wavelength $w$ and link $e$ for its working lightpath, 0 otherwise
			\item $y_{e, w}^{d} \in \{0,1\} $: equals 1 if demand $d$ uses wavelength $w$ and link $e$ for its protection lightpath, 0 otherwise
			\item $\alpha_{w}^{d} \in \{0,1\} $: equals 1 if wavelength $w$ is assigned for demand $d$ on its working lightpath, 0 otherwise
			\item $\beta_{w}^{d} \in \{0,1\} $: equals 1 if wavelength $w$ is assigned for demand $d$ on its protection lightpath, 0 otherwise
			\item $z_{e, w}^{d, v} \in \{0,1\} $: equals 1 if at node $v$, demand $d$ is encoded with another demand and the encoded signal is modulated on wavelength $w$ and routed over link $e$, 0 otherwise
			\item $\theta_{v}^{d} \in \{0, 1\} $: equals 1 if the encoding operation for demand $d$ is taken place at node $v$, 0 otherwise
			\item $f_{d_1}^{d_2} \in \{0, 1\} $: equals 1 if two demands $d_1$ and $d_2$ are encoded with each other, 0 otherwise
			\item $\gamma_{e,w} \in \{0, 1\}$: equals 1 if wavelength $w$ is utilized on link $e$, 0 otherwise 
			\item $\delta_{w} \in \{0, 1\}$: equals 1 if wavelength $w$ is utilized for the entire network, 0 otherwise \\
		\end{itemize}

		\noindent{Objective Function:} 
		
		\begin{equation} \label{eq:obj}
			\textit{Minimize:} \; \; c_1 \times \sum_{w \in W} \delta_{w} + c_2 \times \frac{1}{2} \sum_{d \in D} \sum_{w \in W} |\alpha_{w}^{d} - \beta_{w}^{d}|
		\end{equation}

		\noindent{Subject to the following constraints:}
		
		\begin{equation}\label{eq:c1}
			\sum_{w \in W} {\alpha^d_w} = 1 \; \; \forall d \in D 
		\end{equation}

		\begin{equation}\label{eq:c1x}
			\sum_{w \in W} {\alpha^d_w} = \sum_{w \in W} {\beta^d_w} \; \; \forall d \in D 
		\end{equation}

		\begin{equation} \label{eq:c2}
			\begin{split}
				\sum_{e \in {E}: v\equiv s(e)} {x_{e, w}^{d}}-\sum_{e \in {E}: v \equiv r(e)} {x_{e, w}^{d} }= 		
				\begin{cases} 
					\alpha_{w}^{d} &\mbox{if } v \equiv s(d) \\ 
					-\alpha_{w}^{d}& \mbox{if } v \equiv r(d)\\
					$0$ & otherwise \\
				\end{cases}  \\   \qquad \qquad \forall v \in V, \forall d \in D, \forall w \in W \hfill
			\end{split}
		\end{equation}
		
		\begin{equation} \label{eq:c2x}
			\begin{split}
				\sum_{e \in {E}: v\equiv s(e)} {y_{e, w}^{d}}-\sum_{e \in {E}: v \equiv r(e)} {y_{e, w}^{d} }= 		
				\begin{cases} 
					\beta_{w}^{d} &\mbox{if } v \equiv s(d) \\ 
					-\beta_{w}^{d}& \mbox{if } v \equiv r(d)\\
					$0$ & otherwise \\
				\end{cases}   \\  \qquad \qquad \forall v \in V, \forall d \in D, \forall w \in W \hfill
			\end{split}
		\end{equation}
		
		\begin{align} \label{eq:c3} {
				\sum_{w \in W} x_{e, w}^{d} + \sum_{w \in W} y_{e, w}^{d} \leq 1 \qquad \forall d \in D, \forall e \in E 
			}
		\end{align}
		
		\begin{align} \label{eq:c4} 
			\begin{split}
				\sum_{d \in D} x_{e, w}^{d} + \sum_{d \in D} y_{e, w}^{d} -\frac{1}{2} \sum_{d \in D} \sum_{v \in V} {z_{e, w}^{d, v}} \leq \gamma_{e, w} &  \\
				\qquad \forall e \in E, \forall w \in W
			\end{split}
		\end{align}

		%
		\begin{align} \label{eq:c6} {
				\sum_{v \in V} \theta_{v}^{d} \leq 1 \qquad and \qquad \theta_{v}^{d} = 0 \qquad \mbox{if } v \equiv r(d) \qquad \forall  d \in D
			}
		\end{align}
		
		\begin{align} \label{eq:c6x} {
				\sum_{w \in W} \sum_{e \in E: v \equiv s(e)} z_{e, w}^{d, v} \geq \theta_{v}^{d} \qquad \forall d \in D, \forall v \in V 
			}
		\end{align}
		
		\begin{align} \label{eq:c7} {
				\sum_{d_2 \in D} f_{d_1}^{d_2} \leq 1 \qquad \forall d_1 \in D
			}
		\end{align}
		
		\begin{equation} \label{eq:c8}
			{f^{d_1}_{d_1}} + \sum_{d_2 \in D: r(d_2) \neq r(d_1)}  {f^{d_1}_{d_2}} = 0 \qquad \forall d_1 \in D
		\end{equation}

		\begin{align} \label{eq:c9} {
				f_{d_1}^{d_2} = f_{d_2}^{d_1} \qquad \forall d_1, d_2 \in D
			}
		\end{align}

		\begin{align} \label{eq:c10} {
				\sum_{d_2 \in D} f_{d_1}^{d_2} = \sum_{v \in V} \theta_{v}^{d_1} \qquad \forall d_1 \in D
			}
		\end{align}
		
		\begin{align} \label{eq:c11} {
				\sum_{w \in W} \sum_{v \in V} z_{e, w}^{d_1, v} \leq \sum_{d_2 \in D} f_{d_1}^{d_2} \qquad \forall d_1 \in D, \forall e \in E
			}
		\end{align}
		
		\begin{align} \label{eq:c12} {
				\sum_{w \in W} z_{e, w}^{d, v}  \leq \theta_{v}^{d}   \qquad \forall d \in D, \forall v \in V, \forall e \in E
			}
		\end{align}
		
		\begin{align} \label{eq:c12x} {
				f_{d_1}^{d_2} \leq \frac{1}{2} \sum_{w \in W} \sum_{e \in E} \sum_{v \in V} (z_{e, w}^{d_1, v} + z_{e, w}^{d_2, v}) \qquad
				\forall d_1, d_2 \in D	
			}
		\end{align}
		
		\begin{align} \label{eq:c13} {
				z_{e, w}^{d_1, v} - z_{e, w}^{d_2, v}+f_{d_1}^{d_2} \leq 1 \qquad \forall d_1, d_2 \in D, \forall w \in W, \forall e \in E
			}
		\end{align}
		
		\begin{align} \label{eq:c14} {
				z_{e, w}^{d_2, v} - z_{e, w}^{d_1, v}+f_{d_1}^{d_2} \leq 1 \qquad \forall d_1, d_2 \in D, \forall w \in W, \forall e \in E
			}
		\end{align}
		
		\begin{align} \label{eq:c15} {
				\theta_{v}^{d_1} - \theta_{v}^{d_2}+f_{d_1}^{d_2} \leq 1 \qquad \forall d_1, d_2 \in D, \forall v \in V
			}
		\end{align}
		
		\begin{align} \label{eq:c16} {
				\theta_{v}^{d_2} - \theta_{v}^{d_1}+f_{d_1}^{d_2} \leq 1 \qquad \forall d_1, d_2 \in D, \forall v \in V
			}
		\end{align}
		
		\begin{align} \label{eq:c17} {
				\sum_{w \in W} x_{e, w}^{d_1} + \sum_{w \in W} x_{e, w}^{d_2} + f_{d_1}^{d_2} \leq 2 \qquad \forall d_1, d_2 \in D, \forall e \in E
			}
		\end{align}
		
		\begin{align} \label{eq:c18} {
				\sum_{w \in W} x_{e, w}^{d_1} + \sum_{w \in W} y_{e, w}^{d_2} + f_{d_1}^{d_2} \leq 2 \qquad \forall d_1, d_2 \in D, \forall e \in E
			}
		\end{align}

		\begin{align} \label{eq:c19} {
				\sum_{v \in V} z_{e, w}^{d, v}  \leq y_{e, w}^{d}  \qquad \forall d \in D, \forall e \in E, \forall w \in W
			}
		\end{align}
		
		\begin{equation} \label{eq:c20}
			\begin{split}
				\sum_{w \in {W}} (\sum_{e \in E: i=s(e)} z_{e, w}^{d, v} - \sum_{e \in E: i=r(e)} z_{e, w}^{d, v})=\\
				\begin{cases} 
					\theta_{v}^{d} &\mbox{if } i \equiv v \\ 
					-\theta_{v}^{d} & \mbox{if } i \equiv r(d)\\
					0 & \mbox{otherwise}
				\end{cases} \qquad \qquad \forall d \in D, \forall v \in V, \forall i \in V \hfill
			\end{split}
		\end{equation}
		
		\begin{equation} \label{eq:c21}
			\sum_{e\in E} \gamma_{e,w} \leq |E| \times \delta_{w}  \; \forall e \in E
		\end{equation}	
		
	\end{footnotesize}
	
	The integrated objective in Eq. \ref{eq:obj} composes of two constituents being weighted by coefficients $c_1$ and $c_2$. The first constituent weighted by $c_1$ is to minimize the wavelength count\textemdash the traditional metric in static network planning\textemdash whereas the second one with the coefficient $c_2$ is to minimize the number of client-side connections. Constraints in Eq. \ref{eq:c1} and Eq. \ref{eq:c1x} are to accommodate all traffic demands by finding proper wavelengths for their working and backup signals. Constrains in Eq. \ref{eq:c2} and Eq. \ref{eq:c2x} are traditional flow conservation for both working and protection signal of each demand. The link-disjoint requirement between the working and backup route is defined in Eq. \ref{eq:c3}. The wavelength singleness on each fiber link is imposed by Eq. \ref{eq:c4}. Constraint formulated in Eq. \ref{eq:c6} say that each demand has at most one coding node which must be different from its destination node. That if demand $d$ is encoded at node $v$, there must be a coding link started from node $v$ is captured by Eq. \ref{eq:c6x}. The condition that each demand could be coded with at most one demand sharing the same destination is formulated by Eqs. \ref{eq:c7}, \ref{eq:c8}, and \ref{eq:c9}. Constraints (\ref{eq:c10}), (\ref{eq:c11}), (\ref{eq:c12}) and (\ref{eq:c12x}) are to guarantee if a demand is coded, the respective coding node, coding link(s) and coding wavelength must be found. The fact that if two demands are coded with each other, they must have the same coding node, coding link(s) and coding wavelength are guaranteed by Eqs. \ref{eq:c13}, \ref{eq:c14}, \ref{eq:c15} and \ref{eq:c16}. Constraints (\ref{eq:c17}), (\ref{eq:c18}) impose the recovery condition on two encoded demands, that is, their working path must be link-disjointed and the working path of the first demand must also be link-disjointed to the protection path of the second demand. The recovery condition guarantees that the original signal could be always restored in the event of any singe link failure. The condition that coding only takes place in the protection route and thus, coding link(s) must be on the protection route of each demand is formulated by Eq. \ref{eq:c19}. The conservation of coding flow is ensured by Eq. \ref{eq:c20}. Finally, the definition of using a wavelength for the entire network is brought about by Eq. \ref{eq:c21}. 
	
	\subsection{Linearization Procedure}
	The absolute term in the integrated objective function in Eq. \ref{eq:obj} gives rise to the non-linearity of mathematical model and the solving of such non-linear models are not well-fitted to optimization solvers. This is due to the fact that the state-of-the-art algorithms for non-linear models are not as maturing as for the linear ones. In dealing with this issue, we provide a linearization procedure to transform the original model into the well-established form of integer linear programming.  \\
	
	We introduce a new variable $u^d=\frac{1}{2} \sum_{w \in W} |\alpha_{w}^{d} - \beta_{w}^{d}|$. Note that the value set for $u^d$ is $\{0, 1\}$ in which the lower value is interpreted as the network-side architecture for connection $d$ (i.e., identical wavelength is used for the working and protection signal) while the higher value is hold for client-side implementation. The objective function is thus re-written in the linear form as followed: 
	
	\begin{equation} \label{eq:new-obj-uc}
		c_1 \sum_{w \in W} \delta_{w} + c_2 \sum_{d \in D} u^d
	\end{equation}

	The minimization sense of the objective function and the positive sign for the absolute term in Eq. \ref{eq:obj} lend itself to the adoption of a standard practice for linearizing such absolute term. Specifically, two following auxiliary constraints have to be added as the consequence of introducing the new variable $u^d$ 
	
	\begin{align} \label{eq:c22} {
			u^d \geq \frac{1}{2} \sum_{w \in W} (\alpha_{w}^{d} - \beta_{w}^{d}) \qquad \forall d \in D
		}
	\end{align}
	
	\begin{align} \label{eq:c23} {
			u^d \geq \frac{1}{2} \sum_{w \in W} (\beta_{w}^{d} - \alpha_{w}^{d}) \qquad \forall d \in D
		}
	\end{align}
	
	The integer linear programming model for the RWNCA-PC problem therefore composes of the linearized integrated objective function in Eq. \ref{eq:new-obj-uc}, a set of constraints formulated by equations from Eq. \ref{eq:c1} to Eq. \ref{eq:c21} and the auxiliary constraints in Eqs. \ref{eq:c22} and \ref{eq:c23}.   
	
	\subsection{An Analysis of Weight Coefficients}
	Our mathematical model aims at minimizing the traditional metric\textemdash the number of used wavelengths\textemdash as the prioritized objective and furthermore, minimizing the number of client-side connections as the secondary goal. To reflect such desired priority, weight vectors have to be carefully selected. In this section, we present an analysis on the impact of weight coefficients to the preference of individual objective.\\
	
	For convenience, the first objective is denoted as $O_1$ (i.e., minimize the wavelength numbers) and the second objective is denoted as $O_2$ (i.e., minimize the number of client-side connections which is equivalent to minimize the number of utilized transponders). Note that the bounds for each objective are: $0 \leq O_1 \leq |W| $ and $0 \leq O_2 \leq |D| $ and the smallest variation of each objective is $one$. Three ranges of value for weight coefficients covering all possible cases are considered as following:   \\
	
	\begin{itemize}	
		\item $\frac{c_1}{c_2} > |D|$: Assuming that the first objective has a minimum increase from $O_1$ to $O_1 + 1$) while the second objective is freely varied from $O_2$ to $O_2^{*}$. The net impact on the integrated objective therefore is $\Delta = c_1 \times (O_1 + 1 - O_1) + c_2 \times (O_2^{*}-O_2)$. Given the constraint that $-|D| \leq O_2^{*}- O_2 \leq |D| $ and the case $c_1 > c_2 \times |D|$, the value of $\Delta$ is always greater than zeros and it is implied that the first objective $O_1$ is the dominating term since a smallest increase in the first objective $O_1$ causes an increase in the integrated objective function in spite of any variation of the second objective $O_2$. Similarly, it is straight to reason that a minimum decrease of the first objective $O_1$ will absolutely decrease the overall objective function and this again holds true for an arbitrary change in the second objective $O_2$. This analysis points to the result that the minimization of the integrated objective has to involve minimizing the first objective $O_1$ in the first place and then minimizing the second objective $O_2$. In other words, the objective $O_1$ is strictly prioritized over $O_2$ for the case $\frac{c_1}{c_2} > |D|$. \\
		
		\item $\frac{c_1}{c_2} < \frac{1}{|W|}$: In a close analogy to the above reasoning, it is easy to verify that $O_2$ is strictly prioritized in this range of values for weight coefficients while $O_1$ is the secondary goal. \\
		
		\item $\frac{1}{|W|} \leq \frac{c_1}{c_2} \leq |D|$: Intuitively, it is observed that the relative significance of objective $O_1$ is increased as the value of $\frac{c_1}{c_2}$ is moved toward $|D|$ and conversely, decreased when $\frac{c_1}{c_2}$ is close to $\frac{1}{|W|}$. Likewise, $O_2$ is of greater impact if $\frac{c_1}{c_2}$ is near to $\frac{1}{|W|}$ and of less impact if $\frac{c_1}{c_2}$ is in close proximity to $|D|$. However, without further information, a definitive conclusion on the strict priority of individual objectives could not be mathematically attained for this case. Note that $O_1$ becomes absolutely prioritized if $\frac{c_1}{c_2} > |D|$ while $O_2$ would be exclusively preferred when $\frac{c_1}{c_2} < \frac{1}{|W|}$ \\ 
		
	\end{itemize}
	
	The above analysis provides insights on the determination of weight vectors such that the strict priority of constituent objectives is ensured. For our interest, the weight values satisfying $\frac{c_1}{c_2} > |D|$ will be considered.

	\section{Numerical Evaluations and Discussions}
	\label{sect: result}
\begin{figure}[!ht]
		\centering
		\includegraphics[width=0.7\linewidth, height=5.5cm]{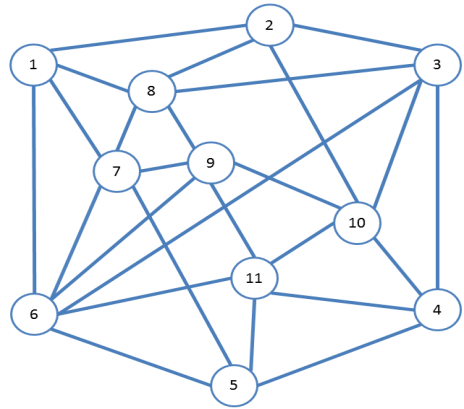}
		\caption{COST239 network}
		\label{fig:cost239}
	\end{figure}

	To quantify the efficiency of our design framework, this section presents numerical evaluations comparing our proposal with other state-of-the-art designs on a realistic COST239 network topology (Fig. \ref{fig:cost239}) whose main characteristics is shown in Tab. \ref{tab:topoinfo}. Note that due to the nature of network coding scheme in our proposal, the selection of network topologies has to satisfy the node degree requirement, that is, greater than or equal to three. We select the COST239 topology for evaluation as it is the widely used topology in many research on optical networks and more importantly, we can have a fair comparison with our previous works \cite{hai_comletter}\textemdash Design 3 in this study. To exploit network coding benefits, we focus on the all-to-one traffic scenario in which a single destination node is chosen and there are requests of a wavelength granularity from the remaining nodes. An extensive benchmark among five designs are performed and the performance comparison is focused on two metrics with priority, that is, the traditional wavelength count is examined first and then among network coding-backed designs, we proceed to compare the second metric, that is, the number of transponders. It is noticed that the number of transponders for each connection is tied to its protection configuration in which the network-side implementation uses one transponder while the client-side scheme requires two transponders. \\
	
	\begin{table}[!ht]
		\caption{Topology Characteristics}
		\label{tab:topoinfo}
		\centering
		\begin{tabular}{|c|c|}
			\hline
			
			\small{$N^o$ Nodes} & \small{11} \\ \hline
			\small{$N^o$ Links} & \small{26x2} \\ \hline
			\small{Nodes of Degree-4} & \small{Node ID: 5, 4, 2, 1} \\ \hline
			\small{Nodes of Degree 5} & \small{Node ID: 11, 10, 9, 8, 7, 3} \\ \hline
			\small{Nodes of Degree 6} & \small{Node ID: 6} \\ \hline
		\end{tabular}
	\end{table}
	
	\begin{table*}[ht]
		\caption{Five Designs under Comparison}
		\label{tab:result0}
		\centering
		\begin{tabular}{ccc}
			\hline
			Design ID & Objective Function & Model Complexity \\
			\hline		
			Design 1 & 1+1 RWA (network-side only): Minimize $\sum_{w \in W} \delta_{w}$ & Variables: $O(|D||E||W|)$; Constraints: $O(|D||W||V|)$   \\
			Design 2 & 1+1 RWA: Minimize $\sum_{w \in W} \delta_{w}$ & Variables: $O(|D||E||W|)$; Constraints: $O(|D||W||V|)$\\
			Design 3 & 1+1 RWNCA (network-side only): Minimize $\sum_{w \in W} \delta_{w}$ & Variables: $O(|D||V||E||W|)$; Constraints: $O(|D^2||E||W|)$ \\
			Design 4 & 1+1 RWNCA: Minimize $\sum_{w \in W} \delta_{w}$  & Variables: $O(|D||V||E||W|)$; Constraints: $O(|D^2||E||W|)$ \\
			Design 5 & 1+1 RWNCA-PC: Minimize $\sum_{w \in W} \delta_{w} + \frac{1}{1+|D|} \sum_{d \in D} u^{d}$  & Variables: $O(|D||V||E||W|)$; Constraints: $O(|D^2||E||W|)$ \\
			\hline
		\end{tabular}  
	\end{table*}

	Table \ref{tab:result0} shows the description of five designs under comparison. The first design (Design 1) is the traditional 1+1 routing and wavelength assignment (RWA) with solely network-side configuration for all connections while in the second design (Design 2), the constraint of same wavelength on the working and protection path is relaxed and thus, such second design could be implemented by a fully client-side mechanism or by a mixed one\textemdash depending on the final solution. That is, if a connection employs different wavelengths for the working and protection, it must be implemented in the client-side configuration whereas if a same wavelength is used for both working and protection path, such connection could be either implemented in the client-side or network-side mode. Clearly, the latter one would be more cost-effective and in this study, we opt for that implementation in such cases. In this context, Design 2 is referred as the mixed protection configuration without NC. The third design (Design 3) is the 1+1 routing, wavelength and network coding assignment (RWNCA) with only network-side protection. The fourth design (Design 4) is the improved version of Design 3 in which the wavelength for working and protection path of a connection could be independently selected. Note that Design 4 is also the network coding-based version of Design 2 in which both network-side and client-side implementation are used while Design 3 is the network coding version of Design 1. Both four designs are the single-objective model aiming at minimizing the number of utilized wavelengths to accommodate all traffic demands. The fifth design (Design 5) is our proposed bi-objective model making use of mixed node architectures to minimize the conventional wavelength number as the primary objective and simultaneously minimize the number of client-side connections as the auxiliary goal. To ensure such order of optimization, we use following weight values $c_1=1$ and $c_2=\frac{1}{1+|D|}$ for the model presented in Section \ref{sect: math}. All five designs based on integer linear programming formulations are solved by CPLEX with academic version and the results are optimally collected for a fair comparison \cite{hai_springer3}. To speed up the running time, advanced implementation techniques have been carefully incorporated including warm-start and distributed parallel optimization. In this study, the most demanding cases have been solved optimally in less than 4 hours which can be acceptable for offline planning. \\
	
	Table \ref{tab:result1} reports the optimal number of used wavelengths to accommodate the all-to-one traffic at all destination node degrees of five designs. As the general trend, the use of photonic XOR in network coding-based designs\textemdash Design 3, 4, 5\textemdash helps to boost the capacity efficiency compared to the conventional non-coding designs\textemdash Design 1, 2. While the use of mixed triggering mechanism with NC\textemdash Design 4 and 5\textemdash consistently surpasses its conventional counterpart (Design 2) at all receiving node degrees, the adoption of NC with network-side triggering features spectral gain only at receiving nodes of degree 5 and 6. This could be owning to the merit of mixed triggering proposal that permits assigning the working and protection wavelength for each demand independently and thus, paving the way for more encoding opportunities among protection flows of demands. With respect to relative enhancement, a highest saving of $25\%$ wavelength resources is noted at node degree 6 for both network-side and mixed mechanism implementation with photonic XOR compared to their respective conventional counterparts. It appears also clear that integrating NC with mixed protection configuration designs\textemdash Design 4 and Design 5\textemdash generally bring about greater performances than integrating NC with the network-side mechanism\textemdash Design 3\textemdash as revealed at node degree 4 and 5 with a relative performance gap of $25\%$ while at node degree 6, both network-side and mixed triggering designs with NC are on the same performance. \\
	
	\begin{table}[!ht]
		\caption{Comparison of Wavelength Count}
		\label{tab:result1}
		\centering
		\begin{tabular}{cccc}
			\hline
			& \multicolumn{3}{c}{Destination Node Type} \\
			\cline{2-4 } 
			& Degree-4 & Degree-5 & Degree-6 \\
			\hline
			Design 1 & 5 & 5 & 4 \\
			Design 2 & 5 & 4 & 4 \\
			\textbf{Design 3} & \textbf{5} & \textbf{4} & \textbf{3} \\
			\textbf{Design 4} & \textbf{4} & \textbf{3} & \textbf{3} \\
			\textbf{Design 5} & \textbf{4} & \textbf{3} & \textbf{3} \\
			\hline
		\end{tabular}
	\end{table}
	
	It has to be noted that Design 4 and Design 5 characterized by the use of mixed protection configuration (i.e., the same wavelength for working and protection signal is relaxed) achieve the highest performance on wavelength efficiency. While Design 4 is the single objective model focusing only on minimizing the wavelength count, Design 5 is the bi-objective model which furthermore includes the minimization of the number of client-side connections. In other words, Design 5 aims to be on a par with Design 4 in terms of wavelength efficiency and yet be more efficient than Design 4 in terms of utilizing transponder resources by optimizing the protection configuration for each connection. \\

		\begin{table*}[!ht]
		\caption{A Comparison on Number of Utilized Transponders for Network Coding-backed Designs}
		\label{tab:result2}
		\centering
		\begin{tabular}{ccccc}
			\hline
			Destination Node & Design ID & \multicolumn{2}{c}{Number of Connection} & Total Number of Transponders \\
			\cline{3-4} 
			& & Client-side & Network-side & \\
			\hline
			
			& Design 3 & 0 & 10 & 10  \\
			Degree 4	& \textbf{Design 4} & \textbf{6} & \textbf{4} & $\textbf{6} \times 2 + \textbf{4}=\textbf{16}$ \\
			& \textbf{Design 5} & \textbf{2} & \textbf{8} & $\textbf{2} \times 2 + \textbf{8}=\textbf{12}$  \\
			\hline
			& Design 3 & 0 & 10 & 10    \\
			Degree 5	& \textbf{Design 4} & \textbf{5} & \textbf{5} & $\textbf{5} \times 2 + \textbf{5}=\textbf{15}$  \\
			& \textbf{Design 5} & \textbf{2} & \textbf{8} & $\textbf{2} \times 2 + \textbf{8}=\textbf{12}$  \\
			\hline
			& Design 3 & 0 & 10 & 10    \\
			Degree 6	& \textbf{Design 4} & \textbf{8} & \textbf{2} & $\textbf{8} \times 2 + \textbf{2}=\textbf{18}$ \\
			& \textbf{Design 5} & \textbf{0} & \textbf{10} & $\textbf{0} \times 2 + \textbf{10}=\textbf{10}$  \\
			\hline
		\end{tabular}
	\end{table*}
	
	For network coding-backed designs\textemdash Design 3, 4, 5\textemdash we proceed to draw a comparison on the number of utilized transponders. Table \ref{tab:result2} shows the protection configuration associated with the optimal solution obtained from each design. Design 3, by definition, makes use of only network-side connections and hence, requires 10 transponders for implementation. Design 4 relaxes the identical wavelength condition of working and protection signal and thus, implies the use of both network-side and client-side implementation. It is shown in Tab. \ref{tab:result2} that Design 4 requires $60\%$ and $50\%$ more transponder resources than Design 3 at receiving node degree $4$ and $5$ respectively and note that in these cases, Design 4 has greater wavelength efficiency than Design 3 (Tab. \ref{tab:result1}). Moreover, at node degree $6$ where Design 3 and Design 4 are on the same performance of wavelength efficiency, Design 4 still consumes $80\%$ more transponders than Design 3. Such excessive use of transponders for Design 4 is due to the fact that Design 4 is focused only on a single objective of wavelength count and thus, the issue of optimizing protection configuration is not taken into account. Design 5 tackles this critical issue with a two-level optimization model in which the wavelength count is optimized first and among the configurations that meet the optimal wavelength count, the most efficient one is selected. As shown in Tab. \ref{tab:result2}, Design 5 represents a transponder saving of $25\%$ and $20\%$ compared to Design 4 at the receiving node degree $4$ and $5$ while achieving as efficient as Design 4 for the wavelength count. Interestingly, at node degree 6, Design 5 consists of only network-side connections, yielding a significant saving of $80\%$ transponders compared to Design 4. Note that at destination node degree 6, the relaxation of identical wavelength constraint between working and protection signal does not produce any gain in wavelength count and therefore, the most efficient configuration is the network-side implementation for the entire network as demonstrated by Design 5. In average, Design 5 features a more than $40\%$ gain in transponder usage in our studied case compared to its single objective counterpart\textemdash Design 4. The merit of Design 5 is therefore on balancing the excessive transponder usage with the wavelength efficiency gain compared to the network-side design. This is thanks to the carefully crafted mathematical model that permits the optimal usage of client-side connections, that is, client-side connection(s) is used if and only if such usage results in an improvement of wavelength efficiency compared to the network-side design.\\    

	\begin{table*}[!ht]
		\caption{Routing, Wavelength Assignment and Dedicated Protection Configuration for Design 4 at Node Degree 5}
		\label{tab:result3}
		\centering
		\begin{tabular}{cccccc}
			\hline
			Demand & W-route & $\lambda_w$ & P-route & $\lambda_p$ & Connection Type\\
			\hline
			1$\rightarrow$3 & (1-6-3) & $\lambda_2$ & (1-2-3) & $\lambda_1$ & \textbf{Client-side}\\
			2$\rightarrow$3 & (2-3) & $\lambda_3$ & (2-10-3) & $\lambda_3$ & Network-side\\
			4$\rightarrow$3 & (4-3) & $\lambda_3$ & (4-10-3) & $\lambda_3$ & Network-side\\
			5$\rightarrow$3 & (5-4-3) & $\lambda_1$ & (5-7-8-3) & $\lambda_3$ & \textbf{Client-side}\\
			6$\rightarrow$3 & (6-3) & $\lambda_3$ & (6-9-8-3) & $\lambda_1$ & \textbf{Client-side}\\
			7$\rightarrow$3 & (7-6-3) & $\lambda_1$ & (7-8-3) & $\lambda_3$ & \textbf{Client-side}\\
			8$\rightarrow$3 & (8-3) & $\lambda_2$ & (8-1-2-3) & $\lambda_1$ & \textbf{Client-side}\\
			9$\rightarrow$3 & (9-10-3) & $\lambda_1$ & (9-8-3) & $\lambda_1$ & Network-side\\
			10$\rightarrow$3 & (10-3) & $\lambda_2$ & (10-2-3) & $\lambda_2$ & Network-side\\
			11$\rightarrow$3 & (11-4-3) & $\lambda_2$ & (11-10-2-3) & $\lambda_2$ & Network-side\\	
			\hline
		\end{tabular}  
	\end{table*}

	\begin{table*}[!ht]
		\caption{Network Coding Configuration for Design 4 at Node Degree 5}
		\label{tab:result4}
		\centering
		\begin{tabular}{cccc}
			\hline
			Coded Demands & Coding Node & Coding links & Coding $\lambda$ \\
			\hline
			$(1 \rightarrow 3) \oplus (8 \rightarrow 3)$ & 1 & (1-2-3) & $\lambda_1$ \\
			$(2 \rightarrow 3) \oplus (4 \rightarrow 3)$ & 10 & (10-3) & $\lambda_3$ \\
			$(5 \rightarrow 3) \oplus (7 \rightarrow 3)$ & 7 & (7-8-3) & $\lambda_3$ \\	
			$(6 \rightarrow 3) \oplus (9 \rightarrow 3)$ & 9 & (9-8-3) & $\lambda_1$ \\	
			$(10 \rightarrow 3) \oplus (11 \rightarrow 3)$ & 10 & (10-2-3) & $\lambda_2$ \\
			\hline
		\end{tabular}  
	\end{table*}

	\begin{table*}[!ht]
		\caption{Routing, Wavelength Assignment and Dedicated Protection  Configuration for Design 5 at Node Degree 5}
		\label{tab:result5}
		\centering
		\begin{tabular}{cccccc}
			\hline
			Demand & W-route & $\lambda_w$ & P-route & $\lambda_p$ & Connection Type\\
			\hline
			1$\rightarrow$3 & (1-2-3) & $\lambda_2$ & (1-6-3) & $\lambda_3$ & \textbf{Client-side}\\
			2$\rightarrow$3 & (2-3) & $\lambda_3$ & (2-10-3) & $\lambda_3$ & Network-side\\
			4$\rightarrow$3 & (4-3) & $\lambda_1$ & (4-10-3) & $\lambda_1$ & Network-side\\
			5$\rightarrow$3 & (5-4-3) & $\lambda_2$ & (5-6-3) & $\lambda_2$ & Network-side\\
			6$\rightarrow$3 & (6-3) & $\lambda_1$ & (6-11-10-3) & $\lambda_1$ & Network-side\\
			7$\rightarrow$3 & (7-8-3) & $\lambda_2$ & (7-6-3) & $\lambda_2$ & Network-side\\
			8$\rightarrow$3 & (8-3) & $\lambda_1$ & (8-2-3) & $\lambda_1$ & Network-side\\
			9$\rightarrow$3 & (9-8-3) & $\lambda_3$ & (9-10-3) & $\lambda_3$ & Network-side\\
			10$\rightarrow$3 & (10-3) & $\lambda_2$ & (10-2-3) & $\lambda_1$ & \textbf{Client-side}\\
			11$\rightarrow$3 & (11-4-3) & $\lambda_3$ & (11-6-3) & $\lambda_3$ & Network-side\\	
			\hline
		\end{tabular}  
	\end{table*}
	
	\begin{table*}[!ht]
		\caption{Network Coding Configuration for Design 5 at Node Degree 5}
		\label{tab:result6}
		\centering
		\begin{tabular}{cccc}
			\hline
			Coded Demands & Coding Node & Coding links & Coding $\lambda$ \\
			\hline
			$(1 \rightarrow 3) \oplus (11 \rightarrow 3)$ & 6 & (6-3) & $\lambda_3$ \\
			$(2 \rightarrow 3) \oplus (9 \rightarrow 3)$ & 10 & (10-3) & $\lambda_3$ \\
			$(4 \rightarrow 3) \oplus (6 \rightarrow 3)$ & 10 & (10-3) & $\lambda_1$ \\	
			$(5 \rightarrow 3) \oplus (7 \rightarrow 3)$ & 6 & (6-3) & $\lambda_2$ \\	
			$(8 \rightarrow 3) \oplus (10 \rightarrow 3)$ & 2 & (2-3) & $\lambda_1$ \\
			\hline
		\end{tabular}  
	\end{table*}
	For completeness of comparison, we present the full simulation results at the receiving node degree 5 (node ID: 3) of Design 4 and Design 5 whose wavelength efficiency are strictly better than other approaches. Note that Design 4 is the single objective version focusing on minimizing the wavelength count while Design 5 is the improved version of Design 4 incorporating two-level optimization, that is, wavelength count first and then the number of client-side connections. The result is to highlight in a more granular manner that given the same wavelength count, the taking into account of protection configuration for optimization as in Design 5 could reduce considerably the redundant resources usage\textemdash number of client-side connections\textemdash compared to Design 4. The full result includes the routing, wavelength assignment for the working and protection signal and furthermore, the protection configuration of each connection as in Tab. \ref{tab:result3} for Design 4 and Tab. \ref{tab:result5} for Design 5. The network coding assignment information is shown in Tab. \ref{tab:result4} and Tab. \ref{tab:result6} for Design 4 and Design 5 respectively. 
	
	\section{Conclusions}
	This paper investigated the routing, wavelength, network coding assignment and protection configuration problem arisen in exploiting photonic XOR network coding for dedicated path protection in transparent WDM networks with a new perspective on the protection triggering mechanism. Different from opaque networks, the triggering configuration in transparent optical networks\textemdash whether it is from client or network side\textemdash leaves major impacts to the amount of redundant resources at the endpoints and consequently to the network design algorithms and hence, to the network performance as the critical issue of wavelength assignment has different constraints with respect to triggering types. To maximize the network performance measured by the wavelength count while minimizing the redundant resources caused by the use of client-side connections, we proposed a mixed use of both network-side and client-side configuration whereas the client-side connection(s) is used only if resulting in a greater network efficiency compared to network-side implementation. A multi-objective optimization model for optimal solving of RWNCA-PC was formulated and developed to minimize the wavelength resources as the prioritized goal and to minimize the number of utilized transponders as the secondary goal. The order of optimization was then examined with a rigorous analysis on the impact of weight coefficients. The efficacy of our design model was evaluated numerically on the realistic network topology, COST239 and extensively benchmarked with reference designs based on single objective optimization models, both with network coding and without network coding. It was revealed that up to $25\%$ saving in wavelength resources could be gained with the application of NC compared to the conventional non-coding designs and designs with mixed protection configuration would be more beneficial from applying NC than the design with solely network-side mechanism. Furthermore, our proposal on a bi-objective model taking advantage of mixed protection configuration outperformed its single-objective counterpart, featuring an average saving of more than $40\%$ transponder usage while simultaneously, achieving the highest spectrum efficiency. Nevertheless, it has to be noted that the precise quantitative conclusions about how much in terms of spectrum and transponder savings that the mixed protection configuration approach could bring about is indeed dependent on the particular network, the traffic type and therefore, there is always a scope for new insights and conclusions when new scenarios are investigated. \\
	
	It has to be noted that the spectral benefits enabled by the use of NC is highly dependent on the underlying network design algorithm and while there have been several extremely efficient heuristic with (nearly-) optimal results for non-coding designs (e.g., routing and wavelength assignment problems), more works are still needed to develop heuristic for coding-aware designs with matched quality. Besides, as more data becomes available, assessing the cost and energy consumption of coding-aware designs remains to be elucidated. In a broader context, the perspective of enabling optical mixing (e.g., photonic network coding) between transitional lightpaths at intermediate nodes paves the way for the arrival of a new architectural paradigm, that is, optical-processing-enabled network which may represent an evolution of the traditional optical-bypass framework. Further works from multiple fronts encompassing enabling technologies to network design algorithms thus need to be carried out to unlock the potential of that new architectural paradigm. \\

	\label{sect: conclusion}
	
	\bibliographystyle{IEEEtran}
	\bibliography{ref} 
	
\end{document}